\documentclass[doublecol]{epl2}
\pdfoutput=1
\usepackage{color}
\usepackage{graphicx,hyperref}% Include figure files
\usepackage{bm}% bold math
\usepackage{marvosym}
\usepackage{times,amsmath,amssymb}
\begin{document}

\title{Mechanism for graphene-based optoelectronic switches by tuning
surface plasmon-polaritons in monolayer graphene}
\shorttitle{Mechanism for graphene-based optoelectronic switches}

\author{Yu. V. Bludov, M. I. Vasilevskiy, and N. M. R. Peres}

\institute{Centro de F\'{\i}sica e Departamento de F\'{\i}sica, Universidade do Minho,
Campus de Gualtar, Braga 4710-057, Portugal}

\abstract{
It is shown that one can explore the optical conductivity of graphene, together
with the ability of controlling its electronic density by an applied gate voltage,
in order to achieve resonant coupling between an external electromagnetic radiation and
surface plasmon-polaritons in the graphene layer.
This opens the possibility of electrical control of
the intensity of light reflected inside a prism placed on top of the graphene layer, by switching between the regimes of total reflection and total absorption. The predicted effect can be used to build graphene-based opto-electronic switches.
}


\maketitle

%-------------------------------------------------------------------------------
% Introduction
%-------------------------------------------------------------------------------

Among the many promised {\it graphene dreams}
\cite{geim,service,epn}, the possibility of exploring the
electronic, thermal, and mechanical properties of graphene,
having in view a new generation of optoelectronic devices, is one of the
most exciting of those {\it dreams}.
Understanding
the fundamental physics of the
interaction of electrons (in graphene) with an electromagnetic
field is a key step toward the realization of such devices.

The optical response of graphene has been an active field of research, both
experimental \cite{kuzmenko,nair,mak,li}
and theoretical \cite{nmrPRB06,falkovsky,carbotte},
and much is already understood. From the theoretical
point of view, the independent electron approximation
predicts that the real part of optical conductivity of graphene, at zero temperature,
has the form $\sigma ^\prime=\sigma_0\theta(\hbar\omega-2\mu)$, where
$\sigma_0=\pi e^2/(2\hbar)$ is the AC universal conductivity of graphene,
$\hbar \omega$ is the photon energy,
$\mu$ is the chemical potential, and $\theta(x)$ is the Heaviside step function.
The imaginary part of the conductivity is finite everywhere, as long as
$\mu$ is also finite \cite{staubergeim}. For zero chemical potential,
the experiments \cite{kuzmenko,nair,mak} confirm the independent electron model
predictions. On the other hand, for finite chemical potential and
$\omega<2\mu$, the experiments show a substantial absorption in this
energy range, at odds with the theoretical prediction. In simple terms,
the real part of the conductivity, as measured experimentally,
follows roughly  the
formula $\sigma ^\prime \simeq\sigma_0\theta(\hbar\omega-2\mu)+0.4 \sigma_0
\theta(2\mu-\hbar\omega)\theta(\hbar\omega-E_D)$, where $E_D$ is the energy
at which the Drude peak starts developing. Below $E_D$, the optical response
increases dramatically.
The difference between experiment and theory can be explained
by both inter-band and intra-band scattering,
due to impurities and electron-electron interactions.
In what concerns our present study, the deviations seen in the experimental data
are actually vital for the effect we discuss below, and therefore the calculations
we present below use the experimentally measured conductivity of graphene.
The exact form of $\sigma (\omega )$ below the $2\mu $ threshold is essential for the particular type
of interaction of the electrons in graphene with an electromagnetic field leading to
the formation of surface plasmon-polaritons \cite{c:SPP-common}.

Surface plasmon-polariton (SPP) is an evanescent electromagnetic wave induced by the coupling of the electromagnetic field to the
electrons near the surface of a metal or a semiconductor. Its amplitude decays exponentially at both sides of
the interface. The SPP properties are determined by the dielectric function of the conductor (which is related to the
optical conductivity) and the dielectric constants of the surrounding media \cite{c:SPP-common,c:SPP-theory}.
Clearly, graphene is the ultimate thin surface layer. This fact, together with the ability of tuning the value
of the chemical potential by an external gate, makes this system particularly
interesting, since the formation of this type of electromagnetic surface waves
can by controlled externally.

%-------------------------------------------------------------------------------
% Fig. 1
%-------------------------------------------------------------------------------
\begin{figure}
\begin{center}
\includegraphics*[width=7cm]{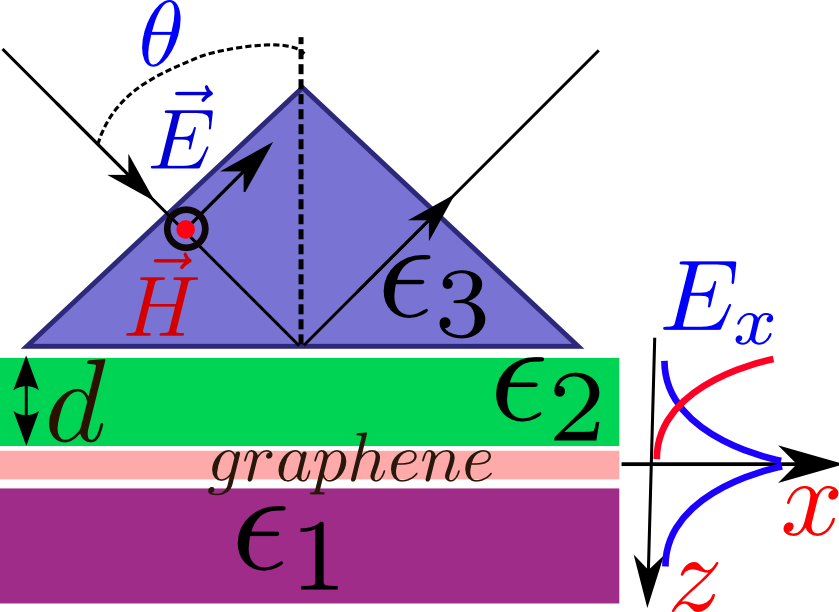}
\end{center}
\caption{Schematic representation of the experimental setup needed to
excite surface plasmon-polaritons in graphene. The graphene layer is sandwiched
between two dielectric media of relative permitivity $\epsilon_1$ (considered semi-infinite) and
$\epsilon_2$ (of width $d$).
On top of the dielectric layer with permitivity
$\epsilon_2$ there is a prism of relative permitivity $\epsilon_3>\epsilon_2$ (which is necessary for total internal reflection). The incident
angle of the incomming light is $\theta$, and the
electric and magnetic fields are
$\bm E=(E_x,0,E_z)$ and $\bm H=(0,H_y,0)$, respectively.
}
\label{fig_plasmon_polaritons}
\end{figure}
%-------------------------------------------------------------------------------
% Fig. 1
%-------------------------------------------------------------------------------

From the point of view of applications, surface plasmon-polaritons can be explored in plasmon sensors \cite{c:sensors} and high-resolution
imaging \cite{c:imaging}, as well as for the miniaturization of photonics components \cite{c:plasmonics1,c:plasmonics2,c:plasmonics3}.
In this Letter, we show that it is feasible to achieve a sharp resonance of the attenuation of an electromagnetic
wave by  transferring its energy into the excitation of surface plasmon-polaritons in graphene.
Indeed, our results demonstrate
that, by adjusing the external gate voltage of an attenuated total internal reflection (ATR) structure with graphene layer, it is
possible to tune, reversibly, the reflectance of the incident electromagnetic wave from total absorption to total reflection.
The structure including a graphene layer, considered in the calculations presented below and suitable for the SPP excitation in graphene is shown in Fig. \ref{fig_plasmon_polaritons} where the necessary definitions are given.

%-------------------------------------------------------------------------------
% Central results
%-------------------------------------------------------------------------------

The geometry of the experimental setup depicted in Fig. \ref{fig_plasmon_polaritons}
is known as ATR configuration.
The key point is to shine light on the prism at an angle larger than the
critical angle ($\theta_c$)
for total internal reflection, characteristic of the interface between
the prism and the dielectric underneath (which can be an air gap or a dielectric layer deposited on top of graphene).
From Snell's law, we have
$\sin\theta_c={\rm max}(\epsilon_1,\epsilon_2)/\epsilon_3$; choosing large $\epsilon_3$,
$\theta_c$ can be made small thus allowing for a broad range of angles, $\theta_c \leq \theta < \pi /2$, to be scanned. Since graphene is at a finite distance
$d$ from the reflecting interface, it is possible to transfer the energy of the incoming
light to the SPP excitation via frustrated (or {\it attenuated}) total
internal reflection.
To compute the electromagnetic response of such a system, we need to solve
Maxwell equations in the presence of the three dielectrics, with graphene acting
as a conductive surface.  The mathematical procedure is well known \cite{c:SPP-theory} but the
calculations are lengthy (we shall outline the derivation procedure below). The central result of this Letter is the equation
defining the reflectance  of an incoming electromagnetic wave using the ATR effect,
with this quantity defined as $R=\vert r\vert^2$ and $r$ being the complex reflection coefficient
given by

\begin{eqnarray}
r=\frac{E^{(r)}_x}{E^{(i)}_x}=
\frac{\left[\epsilon_1/
q_1+4\pi i\sigma/c\right]
\chi_1^- +\epsilon_2\chi_2^-/q_2}
{\left[\epsilon_1/
q_1+4\pi i\sigma/c\right]
\chi_1^+ +\epsilon_2\chi_2^+/q_2}\,,
\label{eq_ramplitude}
\end{eqnarray}
where
\begin{eqnarray}
\chi_1^\pm=\tanh(\kappa q_2d)\pm\frac{q_3\epsilon_2}{q_2\epsilon_3} \,,\\
\chi_2^\pm=1\pm\frac{q_3\epsilon_2}{q_2\epsilon_3}\tanh(\kappa q_2d)\:.
\label{chi}
\end{eqnarray}
In Eq. (\ref{eq_ramplitude}) $q_n=\sqrt{\left (k/\kappa \right )^2-\epsilon_n}$ ($n=1,2,3$) with
\begin{equation}
k=\kappa \sqrt {\epsilon _3}\sin \theta\:,
\label{scanline}
\end{equation}
denoting the in-plane component of the photon wavevector, $\kappa=\omega/c$, and $c$ is the speed of light in vacuum. In what follows we will discuss the
physical effects implied by Eq. (\ref{eq_ramplitude}) and how they can be used to build a graphene-based
optoelectronic switch.

\begin{figure}
\begin{center}
\includegraphics*[width=8.5cm]
{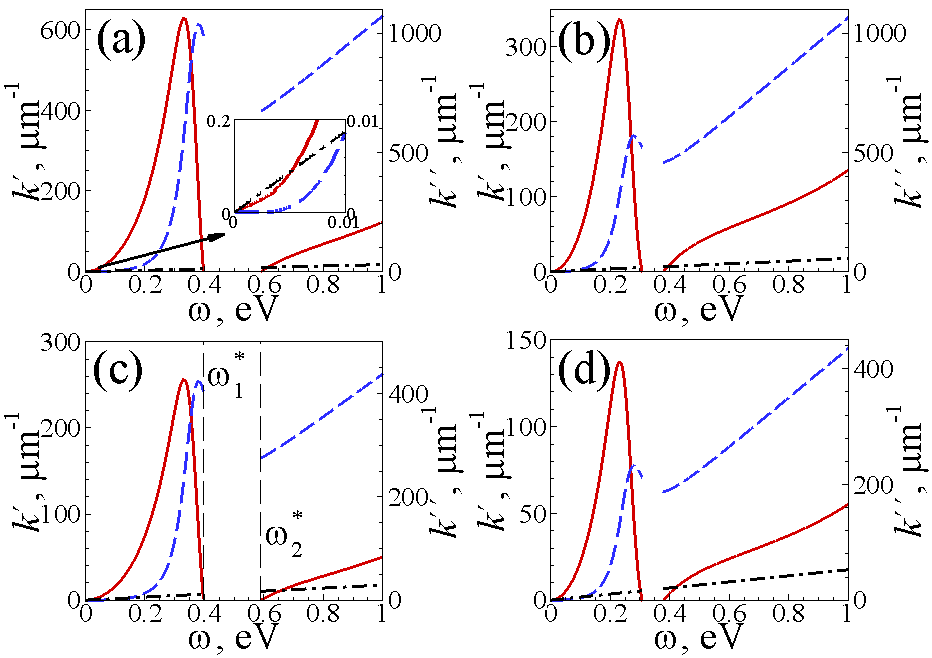}
\end{center}
\caption{ (color on line) Real (solid lines) and imaginary $k^{\prime\prime}$ (dashed lines) parts of the SPP wavevector versus frequency, calculated for a single graphene layer with the following parameters: $\varepsilon_2=1$, $\varepsilon_1=3.9$ (panels a,b), $\varepsilon_1=1$ (panels c,d), $\mu\simeq0.22\,$eV (panels a,c), $\mu\simeq0.16\,$eV (panels b,d). Chemical potentials, corresponding to panels (a) and (b), can be achieved by applying the gate voltage $V=50\,$V or $V=25\,$V, correspondingly, to graphene layer deposited on SiO$_2$ substrate with thickness $300\,$nm \cite{li}. In all panels the ATR scanline for angle of incidence $\Theta=60^\circ$ and prism with $\varepsilon_3=16$ is depicted by the dash-and-dot line. The meaning of critical frequencies $\omega_{1,2}^{*}$ is designated by vertical lines in panel (c).
} \label{fig:dr}
\end{figure}

At resonance between the incident wave and SPP, the amplitude of the reflected wave is minimal and, if the dielectric constants $\epsilon_m$ are all real, the reflectance vanishes, that is $R=0$. At the same time, the SPP amplitude, determined by the field in the medium 2, is given by
\begin{equation}
e^{(2)}_x \equiv \frac{ E^{(2)}_x(0,0)}{E^{(i)}_x}
=\frac{2\epsilon_2D^{-1}(\omega)}{q_2\cosh{(\kappa q_2 d)}}\,
\label{E2}
\end{equation}
with $D(\omega)$ denoting the denominator of Eq. (\ref{eq_ramplitude}).
The SPP-light coupling in graphene can also be understood from the following simplified picture. In the case of $d\rightarrow \infty$, the condition $D(\omega)=0$ reduces to
\begin{equation}
\frac {\epsilon _1} {q_1}+ \frac {\epsilon _2} { q_2}=\frac {4\pi\sigma(\omega)}{ic}\:,
\label{eq:disp-rel}
\end{equation}
which gives the dispersion relation for the SPP excitations in graphene surrounded by two semi-infinite media ($\epsilon_1$ and $\epsilon_2$).
The dispersion curve $\omega(k)$, determined by Eq. (\ref{eq:disp-rel}), allows for a qualitative analysis of the SPP properties and the conditions for their excitation.
We note that in Eq.(\ref{eq:disp-rel}) the wavevector component along the $x$ direction, $k=k^\prime+ik^{\prime\prime}$, is complex and its imaginary part $k^{\prime\prime}$ describes the decay of the electromagnetic wave along the graphene sheet. If there were no dissipation in the graphene layer (i.e. if $\sigma^\prime(\omega)\equiv 0$), SPP would be a nondecaying wave with a purely real wavevector ($k^{\prime\prime}\equiv 0$), propagating along $x$ with a phase velocity $v_f$, smaller than the speed of light in either of the surrounding dielectrics, that is $v_f<c/\sqrt{{\rm max}(\varepsilon_1,\varepsilon_2)}$.
In reality, SPP is decaying because of the energy dissipation in graphene ($\sigma^{\prime}\neq 0$) and the product $k^{\prime} k^{\prime\prime}$ must be positive. It follows from Eq. (\ref{eq:disp-rel}) that the condition  $k^{\prime} k^{\prime\prime}>0$ is satisfied only when $\sigma^{\prime\prime}(\omega)> 0$. \textit{This is a qualitative criterium for the existence of SPPs in the considered structure.} Our numerical results presented in Figs. \ref{fig:dr}(a) to \ref{fig:dr}(d) confirm that for SPPs in graphene the above criterium indeed works.
From these figures we can see that as $\omega$ increases the real part of the wavevector ($k^\prime$) initially increases (hence, SPPs possess a positive group velocity) and then it reaches a maximum and starts decreasing (here the group velocity is negative).
Finally it comes to zero at a critical frequency $\omega^*_1$ where the imaginary part of the conductivity vanishes, $\sigma^{\prime\prime}(\omega^*_1)=0$. Notice that there is a second SPP band at $\omega>\omega^*_2$ (where $\omega_2^*$ is another critical frequency, $\sigma^{\prime\prime}(\omega^*_2)=0$), separated by a finite gap from the lower band
[Figs. \ref{fig:dr}(a) to \ref{fig:dr}(d)]. The gap corresponds to the frequency range $\omega^*_1<\omega<\omega^*_2$ and it is precisely where the imaginary part of the conductivity of graphene is negative and SPPs cannot exist, according to the above criterium.
The imaginary part of the wavevector $k^{\prime\prime}$ increases with the frequency in each of the SPP bands and has its maximum in the vicinity of $\omega^*_1$. By decreasing the chemical potential $\mu$ the maximum values of $k^\prime$ and $k^{\prime\prime}$
also decrease [compare Figs. \ref{fig:dr}(a) and \ref{fig:dr}(b), as well as Figs. \ref{fig:dr}(c) and \ref{fig:dr}(d)]. The same result can be achieved by decreasing the dielectric constants of the surrounding media. For instance, when the graphene layer is deposited on the SiO$_2$ substrate with $\varepsilon_1=3.9$ for the frequency range relevant to our situation [see Figs.\ref{fig:dr}(a),\ref{fig:dr}(b)] the maxima of $k^\prime$ and $k^{\prime\prime}$ are higher than in the case of vacuum [Figs.\ref{fig:dr}(c),\ref{fig:dr}(d)].

Passing to the description of \textit{excitation problem}, this can be seen as a coupling of the exponential tails of the electromagnetic wave in the prism and SPPs in graphene, similar to the tunneling effect. We notice that now the in-plane wavevector of the radiation in the prism is a real value defined by its frequency, the angle of incidence and the dielectric constant of the prism material, as expressed by relation (\ref{scanline}) and known as ATR scan line. The resonant coupling occurs when the ATR scanline crosses the SPP dispersion curve $\omega(k)$ [see inset in Fig. \ref{fig:dr}(a)]. The value of $k^{\prime\prime}$ at the crossing point determines the quality
 factor $Q$
of the resonant excitation and, hence, the relative amplitude of the excited SPP, $e^{(2)}_x$. Consequently, the efficient SPP excitation is only possible at small values of $\omega$, because the large $k^{\prime\prime}$ hampers the excitation in the vicinity of $\omega^*_{1,2}$. Since the imaginary part of the optical conductivity of graphene depends on the gate-controlled chemical potential (for example, at low frequencies, $\omega << |\mu|$, $\sigma^{\prime\prime}\approx 4\sigma _0 |\mu |/(\pi \hbar \omega )$\cite {staubergeim}), it is clear from Eq. (\ref{eq:disp-rel}) that the resonance can be tuned by changing the gate voltage applied to the graphene layer. It should be borne in mind, however, that the true position and the width of the resonance can be found only considering the whole system shown in Fig. \ref{fig_plasmon_polaritons} and including the real part of the optical conductivity of graphene.

\begin{figure}
\begin{center}
\includegraphics*[width=8.5cm]
{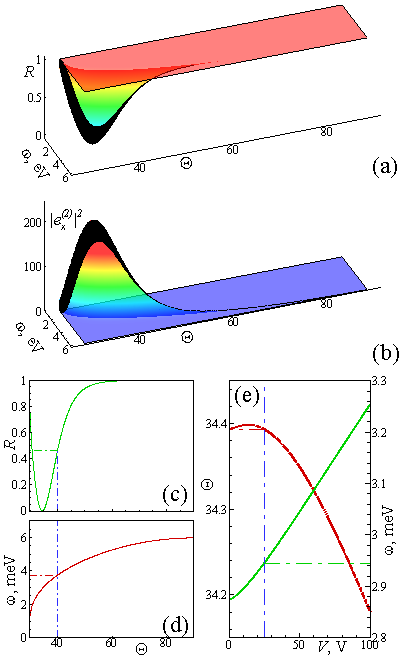}
\end{center}
\caption{(color on line) Panels {\bf (a,b):} reflectivity $R$ [panel (a)] and SPP square amplitude $|e^{(2)}_x|^2$ [panel (b)] \textit{versus} angle of
incidence $\theta$, and frequency $\omega$ for the ATR structure with $\mu\simeq 0.16$ eV
(corresponding to $V=25\,$V of gate voltage); Panel {\bf (c):}
reflectivity $R$ \textit{versus} angle of
incidence $\theta$, corresponding to the minimal value of $R(\omega,\theta)$, depicted in panel (a);
 Panel {\bf (d):}
the frequency $\omega$ at which for given $\Theta$ the minimum of $R$ in panel (a) occurs.
Panel {\bf (e):} angle of incidence,
$\theta$, and frequency, $\omega$, corresponding to zero reflectivity of the ATR
structure \textit{versus} gate voltage $V$. In all panels the parameters of the ATR
structure (Fig. \ref{fig_plasmon_polaritons})
are $\varepsilon_1=3.9$, $\varepsilon_2=1$, $\varepsilon_3=16$,
$d=40\,\mu$m.}
\label{fig:reflectance}
\end{figure}

Figure \ref{fig:reflectance} shows a particular example of ATR via coupling to the gated graphene layer at work.
%-------------------------------------------------------------------------------
% Fig. 2
%-------------------------------------------------------------------------------
%-------------------------------------------------------------------------------
% Fig. 2
%-------------------------------------------------------------------------------
In panels (a)--(d) of this figure, the gate voltage $V$ of the capacitor made of a graphene layer
as one plane of the plates of the field effect transistor
is kept constant. Panel (a) shows the minimum value of $R$  at a particular
value of the pair of variables $\theta$ and  $\omega$ that can be controlled independently. Comparison of panels (a) and (b) demonstrates that resonance minimum of reflectance $R$ corresponds to the maximal amplitude of the excited SPP. In other words, these reflectance minima occur as a result of the transformation of the incident wave energy into the energy of SPPs, excited in graphene.
It is clear that the value of $R$ can be
made very small and even zero by an appropriate choice of $\theta$ and $\omega$ as depicted in panels (b) and (c).
As we move along the curve of panel (c), by changing the angle $\theta$, the value of $R$ at that angle
depends on the frequency of the incoming light. The value of $\omega $, corresponding to the minimum of the reflectivity $R$, can be read off from panel (d) by drawing a straight vertical line
crossing both (c) and (d) panels; an example of such a
line is given in Fig. (\ref{fig:reflectance}). In panel (e) we set $R=0$
and observe how the values of $\theta$ (left axis) and $\omega$ (right axis), change as functions of the gate voltage.
For a given gate voltage, a vertical line is drawn intersecting
both $\theta(V)$ and $\omega(V)$ curves and the pair of these two variables
giving $R=0$ is read off from the corresponding vertical axis.
From this panel, we see that it is always possible to find a pair of
$\theta(V)$ and $\omega(V)$ satisfying the condition for total absorption.
Of course, it is possible to draw the corresponding curves for any
given value of $R$. Thus, we can see that indeed it is possible to
tune externally the resonance absorption condition by changing the external gate voltage.
We would like to stress at this point that it is feasible to implement
the predicted resonance absorption effect using large area graphene samples grown by chemical vapor
deposition (CVD) \cite{kim,rod},
since also in this case the optical conductivity of graphene does not change substantially
from that of exfoliated produced samples. Since the CVD method
allows graphene to be transferred to different substrates, it is also feasible to realize the ATR structure depicted in Fig. \ref{fig_plasmon_polaritons}.

%-------------------------------------------------------------------------------
% End of central results
%-------------------------------------------------------------------------------

\begin{figure}
\begin{center}
\includegraphics*[width=8.5cm]{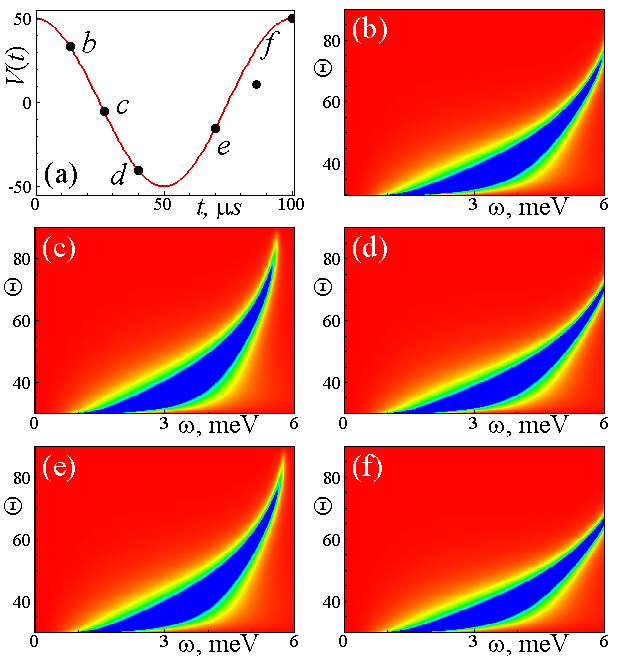}
\end{center}
\caption{ Panels {\bf (a)}: shape of the ac gate voltage $V(t)$ with amplitude $50\,$V and frequency $10\,$kHz. Panels {\bf (b-f)}: reflectance $R(\Theta,\omega)$ for the attenuated total reflection structure at time moments, corresponding to points b--f at panel (a), is depicted in respective panels. The parameters of the ATR structure are the same as in Fig.\ref{fig:reflectance}.}
\label{fig:alt-curr}
\end{figure}

%-------------------------------------------------------------------------------
% Detailed analysis
%-------------------------------------------------------------------------------

%-------------------------------------------------------------------------------
% Fig. 3
%-------------------------------------------------------------------------------
\begin{figure}
\begin{center}
\includegraphics*[width=8.5cm]
{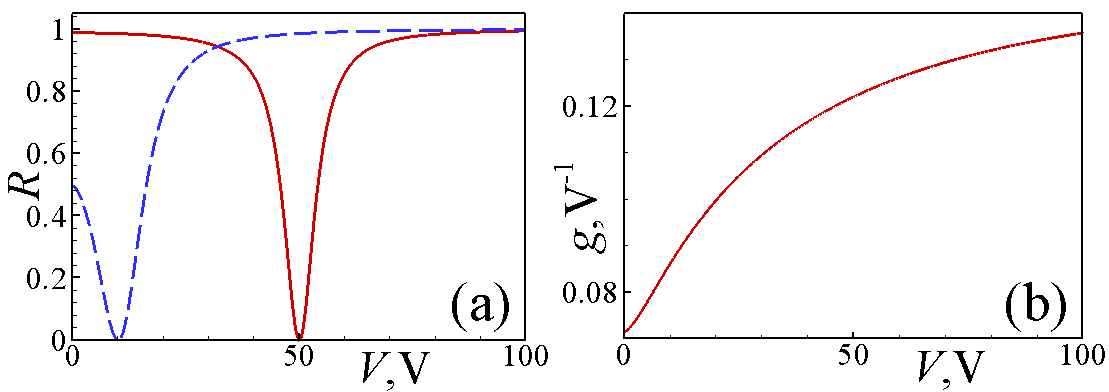}
\end{center}
\caption{(color on line) Panel {\bf (a):} reflectivity $R$ \textit{versus} gate voltage $V$ for ATR structure with $\Theta=34.40\,^\circ$, $\omega=2.894\,$meV [dashed line] or $\Theta=34.35^\circ$, $\omega=3.044\,$meV [solid line]; Panel {\bf (b):}
the abruptness $g=dR/dV$ as function of gate voltage $V$ of the characteristic $R(V)$, calculated on the level $R=0.5$ for $\Theta$, $\omega$, corresponding to those in Fig.\ref{fig:reflectance}(e). Other parameters of the ATR structure are the same as in Fig. \ref{fig:reflectance}.}
\label{fig:rv}
\end{figure}
%-------------------------------------------------------------------------------
% Fig. 3
%-------------------------------------------------------------------------------

The possibility of changing the chemical potential in graphene  by applying a gate voltage makes it possible to tune the ATR conditions by changing this latter quantity, similar to the tuning of the plasmon spectrum as proposed in Ref. \cite{c:graphene-plas-tuning}. Fig.\ref{fig:alt-curr} demonstrates the variation in time of the reflectivity of the ATR structure as the function of frequency and angle of incidence. The gate voltage is changed adiabatically according to the harmonic law $V(t)=50\cos(2\pi ft)$, with the frequency $f=10\,$kHz [Fig.\ref{fig:alt-curr}(a)]. As it can be seen from Fig.\ref{fig:alt-curr}, a decrease of the absolute value of gate voltage corresponds to a shift of the polaritonic resonance to the low-frequency region [e.g., compare Fig.\ref{fig:alt-curr}(b) and \ref{fig:alt-curr}(c), as well as Fig.\ref{fig:alt-curr}(d) and \ref{fig:alt-curr}(e)] and a broadening of the resonance.

Now a natural question arises: how sensitive is the reflectance of the ATR structure to the variation of gate voltage, when frequency $\omega$ and angle of incidence $\Theta$ are fixed? The answer follows from Fig. \ref{fig:rv}(a), which demonstrates the possibility of changing the ATR structure's reflectivity from zero to almost unity (full reflectance) by varying the gate voltage within a range of $\delta V\sim 10\,$V. As a result, the ATR structure with graphene layer can operate as \textit{a switch of electromagnetic radiation} in the \textit{terahertz} domain, where the intensity of the reflected electromagnetic wave can be tuned by the gate voltage, applied to graphene. An important characteristic of the THz switch is the sharpness of the resonance, $g$, which is analogous to the cross-conductivity of a field-effect transistor, calculated as the derivative $g=dR/dV$ at the level $R=0.5$. As one can see from Fig. \ref{fig:rv}(b), $g$ shows an increase with $V$ and then saturates at $V\approx 100\,$V.

We shall now present the derivation of our main result. The solution of Maxwell equations ${\rm rot}{\vec{E}^{(m)}}=i\kappa\vec{H}^{(m)}$, ${\rm rot}{\vec{H}^{(m)}}=-i\kappa\varepsilon_m\vec{E}^{(m)}$ for the $p$-polarized wave can be written in the form:
\begin{eqnarray}
E^{(m)}_x(x,z)&=&\left[A_+^{(m)}\exp(\kappa q_m z)+A_-^{(m)}\exp(-\kappa q_m z)\right]e^{ikx}, \label{eq:sol1}\nonumber\\
H^{(m)}_y(x,z)&=&-\frac{\kappa\varepsilon_m}{k}E^{(m)}_z(x,z)
=
\frac{i\varepsilon_m}{q_m}\left[A_+^{(m)}\exp(\kappa q_m z)\right.\nonumber\\
&&\left.-A_-^{(m)}\exp(-\kappa q_m z)\right]e^{ikx}\nonumber.
\end{eqnarray}
The meaning of the coefficients $A_\pm^{(m)}$ is different for the different media. Since in the medium 3 the electromagnetic wave is propagating in $z$-direction, $q_3$ is imaginary [${\rm Im}(q_3)<0$], and the coefficients $A_+^{(3)}=E^{(r)}_x\exp(\kappa q_3d)$, $A_-^{(3)}=E^{(i)}_x\exp(-\kappa q_3d)$ are proportional to the amplitudes (at $z=-d$) of the reflected $E^{(r)}_x$ and incident $E^{(i)}_x$ waves, respectively. In the media 2 and 1 the electromagnetic waves are exponentially decaying along the $z$-axis, and $q_m$ are real ($q_2,\:q_1>0$). In the medium 2, the coefficients $A_\pm^{(2)}=\left[E^{(2)}_x(0,0)\pm(q_2/i\varepsilon_2)H^{(2)}_x(0,0)\right]/2$ are superpositions of the electric $E^{(2)}_x(0,0)$ and magnetic $H^{(2)}_y(0,0)$ fields at boundary $z=0$. In the medium 1, the coefficient $A_+^{(1)}=0$, while $A_-^{(1)}=E^{(1)}_x(0,0)$ stands for the electric field amplitude at $z=0$. Boundary conditions at $z=-d$ imply the continuity of the tangential components of the electric and magnetic fields,  [$E_x^{(3)}(x,-d)=E_x^{(2)}(x,-d)$, $H_y^{(3)}(x,-d)=H_y^{(2)}(x,-d)$]. At $z=0$ the tangential component of the electric field is continuous $E^{(1)}_x(x,0)=E^{(2)}_x(x,0)$, while the discontinuity of the tangential component of the magnetic field,  $H^{(1)}_y(x,0)-H^{(2)}_y(x,0)=-(4\pi/c)j_x=-(4\pi/c)\sigma(\omega)E_x(x,0)$, stems from the presence of surface currents (caused by the SPP electric field) in the graphene layer. Matching these boundary conditions, we obtain a system of linear equations for the field amplitudes in the matrix form,
\begin{eqnarray}
\left(
\begin{array}{ccc}
C_h & \frac{iq_2}{\varepsilon_2}S_h & -1 \\
\frac{\varepsilon_2q_3}{\varepsilon_3q_2}S_h & \frac{iq_3}{\varepsilon_3}C_h & 1 \\
\frac{i\varepsilon_1}{q_1}-\frac{4\pi}{c}\sigma & 1 & 0
\end{array}\right)
\left(
\begin{array}{c}
e^{(2)}_x \\ h^{(2)}_y \\ r
\end{array}
\right)=
\left(
\begin{array}{c}
1 \\ 1\\ 0
\end{array}
\right).
\label{matrix}
\end{eqnarray}
where $C_h=\cosh(\kappa q_2d)$, $S_h=\sinh(\kappa q_2d)$, and $h^{(2)}_y=H^{(2)}_y(0,0)/E^{(i)}_x$. Solution of (\ref{matrix}) yields the reflection coefficient [Eq. (\ref{eq_ramplitude})] and the field amplitude in the gap [Eq. (\ref{E2})].

To conclude, we have demonstrated that the ATR structure incorporating a monolayer graphene sheet can operate as THz switch where the reflectance of an electromagnetic wave can be switched from nearly unity to nearly zero by applying an external gate voltage to the graphene layer. Since the typical frequencies are $\sim 5\,$meV (or $\sim 1.2\,$THz, which corresponds to $\sim 0.25\,$mm of wavelength in vacuum), this structure can operate in the submillimeter range. The frequency of excited SPPs can be increased by using a prism material with a higher dielectric permeability $\epsilon_3$. The proposed device can also be used for spectroscopy of the graphene optical
 conductivity \cite{PERES} through measuring the characteristics of the ATR-excited SPPs.

%-------------------------------------------------------------------------------
% End of the Detailed analysis
%-------------------------------------------------------------------------------

%-------------------------------------------------------------------------------
% References
%-------------------------------------------------------------------------------

%-------------------------------------------------------------------------------
% End document
%-------------------------------------------------------------------------------

\begin{thebibliography}{0}
\bibitem{geim}
    \Name{Geim A. K.}
    \REVIEW{Science}{324}{2009}{1530}.

\bibitem{service}
    \Name{Service R. F.}
    \REVIEW{Science}{324}{2009}{877}.

\bibitem{epn}
    \Name{Peres N. M. R.}
    \REVIEW{Europhysics News}{40}{2009}{17}.

\bibitem{kuzmenko}
    \Name{Kuzmenko A. B. \and {\it et al.}}
    \REVIEW{Phys. Rev. Lett.}{100}{2008}{117401}.


\bibitem{mak}
    \Name{Kin Fai Mak  \and {\it et al.}}
    \REVIEW{Phys. Rev. Lett.}{101}{2008}{196405}.

\bibitem{li}
    \Name{Li Z. Q. \and {\it et al.}}
    \REVIEW{Nature Phys.}{4}{2008}{532}.

\bibitem{nair}
    \Name{Nair R. R. \and {\it et al.}}
    \REVIEW{Science}{320}{2008}{1308}.

\bibitem{nmrPRB06}
    \Name{Peres N. M. R. \and {\it et al.}}
    \REVIEW{Phys. Rev. B}{73}{2006}{125411}.

\bibitem{falkovsky}
    \Name{Falkovsky L. A. \and Pershoguba S. S.}
    \REVIEW{Phys. Rev. B}{76}{2007}{153410}.

\bibitem{carbotte}
\Name{Gusynin V. P. \and {\it et al.}}
\REVIEW{New J. Phys.}{11}{2009}{095013}.

\bibitem{staubergeim}
\Name{Stauber T. \and {\it et al.}}
\REVIEW{Phys. Rev. B}{78}{2008}{085432}.

\bibitem{c:SPP-common}
 H. R\"ather, {\it
Surface Plasmons on Smooth and Rough Surfaces and on Gratings},
(Springer-Verlag: Berlin, 1988).

\bibitem{c:SPP-theory}
    \Name{Pitarke J. M. \and Silkin V. M. \and Chulkov E. V. \and Echenique P. M.}
    \REVIEW{Rep. Prog. Phys.}{70}{2007}{1}.

\bibitem{c:sensors}
    \Name{Homola J. \and {\it et al.}}
    \REVIEW{Sensors and Actuators B}{54}{1999}{3}.

\bibitem{c:imaging}
    \Name{Kawata S. \and {\it et al.}}
    \REVIEW{Nature Photonics}{3}{2009}{388}.

\bibitem{c:plasmonics1}
    \Name{Vasa P. \and \textit{et al.}}
    \REVIEW{Laser \& Photon. Rev.}{3}{2009}{483}.

\bibitem{c:plasmonics2}
    \Name{Zayats A. V. \and {\it et al.}}
    \REVIEW{Physics Reports}{408}{2005}{131}.

\bibitem{c:plasmonics3}
    \Name{Barnes  W. L. \and {\it et al.}}
    \REVIEW{Nature}{424}{2003}{824}.

\bibitem{kim}
    \Name{Keun Soo Kim \and {\it et al.}}
    \REVIEW{Nature}{457}{2009}{706}.

\bibitem{rod}
    \Name{Xuesong Li \and {\it et al.}}
    \REVIEW{Science}{324}{1312}.

\bibitem{c:graphene-plas-tuning}
    \Name{Ryzhii V. \and Satou A. \and Otsuji T.}
    \REVIEW{J. Appl. Phys.}{101}{2007}{024509}.

\bibitem{PERES}
\Name{Peres N.M.R.}
    \REVIEW{Rev. Mod. Phys}{82}{2010}{2673}.

%-------------------------------------------------------------------------------
% END References
%-------------------------------------------------------------------------------
\end{thebibliography}
\end{document}